\documentclass[11pt]{article}
\usepackage{cite}
\usepackage{amsmath,amsfonts,amssymb}
\usepackage{graphicx,color}
\usepackage{epsfig,epsf}
\usepackage{bm}

\usepackage{relsize}
\def\Babar{{\mbox{\slshape B\kern-0.1em{\smaller A}\kern-0.1em B\kern-0.1em{\smaller A\kern-0.2em R}}}}

\usepackage[bookmarks, breaklinks, colorlinks,urlcolor=black, citecolor=red, 
linkcolor=blue]{hyperref}

\usepackage{url}

 \usepackage{color}

 \usepackage[normalem]{ulem}

 \definecolor{darkgreen}{cmyk}{1,0,1,0.4}
 
 \definecolor{pink}{cmyk}{0.4,1,0.3,0}

\def\com2#1{\textcolor{red}{\it{#1}}}

\def\dsp{\displaystyle}

\def\g {\gamma}

\def\bar {\overline}

\def\bea {\begin{eqnarray}}
\def\eea {\end{eqnarray}}

\def\rd {R(D)}
\def\rdst {R(D^*)}
\def\rdrdst {R(D^{(*)})}

\def\rk {R_K}
\def\rkst {R_{K^*}}
\def\rkrkst {R_{K^{(*)}}}
\def\rJpsi{R_{J/\psi}}

\def\st {\sin\theta}
\def\ct {\cos\theta}

\def\g{\gamma}

\def\beq{\begin{equation}}
\def\eeq{\end{equation}}
\def\barr{\begin{array}}
\def\earr{\end{array}}
\def\dis{\displaystyle}
\def\ltap{\raisebox{-.4ex}{\rlap{$\sim$}} \raisebox{.4ex}{$<$}}

\def \azeL{{{\cal A}_0^L}}
\def \azeR{{{\cal A}_0^R}}
\def \apaL{{{\cal A}_\parallel^L}}
\def \apaR{{{\cal A}_\parallel^R}}
\def \apeL{{{\cal A}_\perp^L}}
\def \apeR{{{\cal A}_\perp^R}}
\def \Re{\text{Re}}

\def\abs#1{\left| #1 \right|}

\def\gev{\ensuremath{\mathrm{Ge\kern -0.1em V}}}
\def\Cn{\ensuremath{C^\mathrm{NP}\,}}
\def\Cten{\ensuremath{C_{10}^\mathrm{NP}\,}}
\def\Cnu{\ensuremath{C_\nu^\mathrm{NP}\,}}
\def\Cnine{\ensuremath{C_{9}^\mathrm{NP}\,}}

\begin{document}

\renewcommand*{\thefootnote}{\fnsymbol{footnote}}

\begin{center}
 {\Large\bf{$\bm {R_{K^{(*)}}}$ and $\bm {R(D^{(*)})}$ anomalies resolved with lepton mixing}}
 \\[6mm]
 {Debajyoti Choudhury $^1$\footnote{Electronic address: debajyoti.choudhury@gmail.com}, 
 Anirban Kundu $^2$\footnote{Electronic address: akphy@caluniv.ac.in}, 
 Rusa Mandal $^3$\footnote{Electronic address: rusam@imsc.res.in}
and Rahul Sinha $^3$}\footnote{Electronic address: sinha@imsc.res.in}
\\[3mm]

$^1${\small\em Department of
  Physics and Astrophysics, University of Delhi, Delhi 110007, India}\\
$^2${\small\em Department of Physics, University of Calcutta,
92 Acharya Prafulla Chandra Road, Kolkata 700009, India}\\
$^3${\small\em The Institute of Mathematical Sciences, HBNI, Taramani, Chennai 600113, India}
 \end{center}


\begin{abstract}

In a recent paper \cite{our-prl}, we had advanced a
minimal resolution of some of the persistent anomalies in semileptonic
$B$-decays. These include the neutral-current observables $R_K$ and
$R_{K^*}$, as well as the charged-current observables $R(D)$ and
$R(D^*)$. Recently, it has been observed that the semileptonic decays
of the $B_c$ meson also hint at a similar type of anomaly. In this
longer version, we discuss in detail why, if the anomalies are indeed
there, it is a challenging task to explain the data consistently in
terms of a simple and compelling new physics scenario.  We find that
the minimal scheme to achieve a reasonable fit involves the inclusion
of just two (or, at worst, three with a possible symmetry relationship
between their Wilson coefficients) new current-current operators,
constructed in terms of the flavour eigenstates, augmented by a change
of basis for the charged lepton fields. With only three unknown
parameters, this class of models not only explain all the anomalies
(including that in $B_c \to J/\psi \, \ell \nu$) to a satisfactory
level but also predict some interesting signatures, like $B\to
K\mu\tau$, $B_s\to \tau\tau$, $B\to K$ plus missing energy, or direct
production of $\tau^+\tau^-$, that can be observed at LHCb or
Belle-II.

\end{abstract}



\setcounter{footnote}{0}
\renewcommand*{\thefootnote}{\arabic{footnote}}


\section{Introduction}

The last few years have seen some intriguing hints of discrepancies in
a few charged- as well as neutral-current decays of $B$-mesons, when
compared to the expectations within the Standard Model (SM). While the
fully hadronic decay modes are subject to large (and, in cases,
not-so-well understood) strong interaction corrections, the situation
is much more under control for semileptonic decays, where the dominant
uncertainties come from the form factors and quark mass values.  Even
these uncertainties are removed to a large extent if one considers
ratios of similar observables.  Thus, the modes $b\to c
\ell\bar\nu$ and $b\to s\ell^+ \ell^-$ are of great
interest. While, individually, none of these immediately calls for the
inclusion of New Physics (NP) effects (given the current significance
levels of the discrepancies), viewed together, they strongly suggest
that some NP is lurking around the corner (see, {\em e.g.},
Ref.~\cite{rusa_sinha}).  Most interestingly, the pattern also argues
convincingly for some NP that violates lepton-flavour universality
(LFU), a cornerstone of the SM. The violation
is to a level that cannot be
simply explained by the inclusion of right-handed neutrino fields and
the consequent neutrino masses; indeed, the strength is only 
a little over an order of magnitude below that generated from the weak scale dynamics.

As we have just mentioned, relative partial widths (or, equivalently,
the ratios of branching ratios (BR)) are particularly clean probes of
physics beyond the SM, on account of the cancellation of the leading
uncertainties inherent in individual BR predictions. Of particular
interest to us are the ratios $R(D)$ and $R(D^*)$ pertaining to
charged-current decays, defined as
\beq
\label{eq:RD}
\rdrdst \equiv \frac{ {\rm BR}(B\to D^{(*)}\tau\nu)}{ {\rm BR}(B\to D^{(*)}\ell \nu)}\,,
\eeq
with $\ell=e$ or $\mu$, the ratio $R_{J/\psi}$ defined as
\beq
\label{eq:defrjpsi}
\rJpsi \equiv \frac{ {\rm BR}(B_c\to J/\psi\, \tau\nu)}{ {\rm BR}(B_c\to J/\psi\, \mu \nu)}\,,
\eeq
and analogous ratios
 for the neutral-current sector
\beq
\label{eq:RK}
\rkrkst \equiv \frac{ {\rm BR}(B\to K^{(*)} \mu^+\mu^-)}{ {\rm BR}(B\to K^{(*)}e^+ e^-)}\,.
\eeq
For the $K^{*}$ mode, the discrepancy is visible not only in the
ratios of binned differential distribution for muon and electron mode,
but also in some angular asymmetries in $B\to K^*\mu\mu$ which we
discuss later.

The SM estimates for these decays are already quite robust. With the
major source of uncertainty being the form factors, they cancel out in
ratios like $\rdrdst$, $\rJpsi$, or $\rkrkst$.  The values of $R(D)$ and
$R(D^{*})$ as measured by \Babar~\cite{Lees:2013uzd}, when taken
together, exceed SM expectations by more than $3 \sigma$.  While the
Belle measurements lie in between the SM expectations and the
\Babar~measurements and are consistent with both
\cite{Huschle:2015rga}, their result on $R(D^{*})$
\cite{Abdesselam:2016cgx}, with the $\tau$ decaying semileptonically,
agrees with the SM expectations only at the $1.6 \sigma$
level\footnote{On the other hand, the measurement of
  $\tau$-polarisation for the decay $B\to D^*\tau\nu$ in Belle
  \cite{Hirose:2016wfn} is consistent with the SM predictions, albeit
  with only a large uncertainty.}.  Similarly, the first measurement
by LHCb \cite{Aaij:2015yra} is also $2.1 \sigma$ above the SM
prediction.  Considering the myriad results together, and including
the correlations, the tension between data and SM is at the level of
$4.1\sigma$ \cite{hfag}.

While the data on $\rd$ and $\rdst$ lie above the SM predictions, 
those on $\rk$ and $\rkst$ are systematically below
the expectations. A similar shortfall has been observed in the $q^2\in [1:6]$ GeV$^2$ bin 
for the decay $B_s\to \phi\mu\mu$,
which is again mediated by the process $b\to s\mu\mu$.  However,
no appreciable discrepancy is found between the data on the
purely leptonic decay $B_s\to\mu^+\mu^-$ and the radiative decay $B\to
X_s\gamma$, and the corresponding SM expectations. The same is true
for the mass difference $\Delta M_s$ and mixing phase $\phi_s$
measurements for the $B_s$ system.  The pattern of deviations is thus
a complicated one and, naively at least, seemingly contradictory.
These, therefore, pose an interesting challenge to the model builders:
how to incorporate all the anomalies in a model with the least number
of free parameters?

While both model-dependent and model-independent search strategies for
NP based on $\rk$ and/or $\rdrdst$ data are being discussed in the
literature for a while now~\cite{oldlit,byakti}, the subject has recently
received a further impetus from the announcement of the apparent
deficit in $\rkst$ \cite{Aaij:2017vbb}. There has been a flurry of activity
trying to explain the $\rk$ and $\rkst$ anomalies within the context
of such models, or in a model-independent framework \cite{rknew}.
However, only a handful of them try to explain both $\rkrkst$ and
$\rdrdst$ anomalies together \cite{newlit}.  Unfortunately, either the
models are not minimal in nature, or the fits are not very good.
Following our earlier rather concise work \cite{our-prl} of a
bottom-up model-independent explanation of both the anomalies, we will
explain, in this paper, the framework in detail, and also expand the
earlier study significantly.

Rather than advocating a particular model, we shall assume a very
phenomenological approach. Virtually all the data can be explained in
terms of an effective Lagrangian, operative at low-energies, and
different from that obtained within the SM. However, with `minimality'
being a criterion, we would like to follow the principle of Occam's
razor and not introduce an arbitrarily large number of new parameters,
as would be the case with a truly anarchic effective theory.
The issue of operator mixing and possible generation of new operators at the scale $m_b$ is
also nontrivial, as shown in Ref.\ \cite{feruglio}. Rather
  than considering the complete set of dimension-6 current-current
  operators and obtaining the best fit to all data, our approach could
  be deemed as an attempt for an inspired guess for the minimal set of operators that
  still produces a more than satisfactory fit, yet with small values
  of the Wilson coefficients (WC). If the NP scale is not beyond a
  few TeVs, the aforementioned operator mixing would be small and the
  WCs of any new operator generated as a quantum corrections would be tiny.
The best fit for the WCs, thus obtained, would presumably pave the way for inspired
model building. We will show how the relationship between the NP WCs
may hold the key for a yet unknown flavour dynamics.

As the reader would have noticed, we have considered only the operators built out of vector and axial-vector currents alone. This is solely because our aim has been  to correlate the charged-current and the neutral-current anomalies and demonstrate that they may have originated from the same source. While the charged-current anomalies in $R(D)$ and $R(D^*)$ can possibly be addressed with scalar and/or tensor current operators while evading the strong constraints arising from the lifetime of the $B_c$ meson \cite{bc-grinstein}, such operators are not of much help for $\rkrkst$. For example, it has widely been discussed in the literature that if the only new operators are those constructed with (pseudo)scalar currents, the explanation for $R_K$ would be incompatible with the data on $B_s\to \ell^+\ell^-$. Similarly, the viability of tensor operators in the case of the neutral current anomalies is discussed in Ref.\ \cite{byakti}. Once one deviates from the assumption of minimality, it is indeed possible to have different Lorentz structures of the NP operators. While this may actually occur in some specific NP models, in the absence of a well-motivated ultraviolet-complete theory, such an Ansatz would militate against the spirit of Occam's razor. 

The rest of the paper is arranged as follows. In Sec.~\ref{sec:data},
we recount and discuss the experimental situation. This is followed,
in Sec.\ \ref{sec:SM}, by a discussion of the microscopic dynamics
that lead to such processes. Sec.\ \ref{sec:model} discusses a model that could have explained the
data but falls short narrowly. This is followed, in
Sec.~\ref{sec:model_new}, by a discussion of more realistic scenarios.
Sec.\ \ref{sec:results} contains our results, and finally we conclude
in Sec.\ \ref{sec:conclusion}.


\section{The data : a brief recounting}
\label{sec:data}

We begin by briefly reviewing the experimental measurements and
theoretical predictions for the observables of interest. We also take
this opportunity to review some further processes that would turn out
to have important consequences in our attempt to explain the
anomalies.

\begin{itemize}

\item
As already mentioned, discrepancies in the measurements of the
observables $\rd$ and $\rdst$ have been seen, over the last several
years, in multiple experiments such as LHCb, \Babar~ and Belle.  In
Table.~\ref{tab:RDRDst}, taken from Ref.\cite{Choudhury:2016ulr}, we
summarize the measurements along with the corresponding SM
predictions.

\begin{table}[!hbt]
\begin{center}
\renewcommand{\arraystretch}{1.2}
\begin{tabular}{lll}
\hline
\hline
        & ${R}(D)$                      & ${R}(D^*)$ \\
\hline
\noalign{\vskip1pt}
SM  prediction   & $0.300 \pm 0.008$ \cite{Na:2015kha} &  $0.252 \pm 0.003$ \cite{Kamenik:2008tj}\\
\Babar~  (Isospin constrained)  & $0.440 \pm 0.058 \pm 0.042 $ & $0.332 \pm 0.024 \pm 0.018 $ \cite{Lees:2013uzd}\\
Belle (2015)  & $0.375 \pm 0.064 \pm 0.026 $ & $0.293 \pm 0.038 \pm 0.015 $ \cite{Huschle:2015rga}\\
Belle (2016)  & --- & $0.302 \pm 0.030 \pm 0.011 $ \cite{Abdesselam:2016cgx}\\
Belle (2016, Full)  & ---& $0.270 \pm 0.035 ~^{+ 0.028}_{-0.025} $ \cite{Hirose:2016wfn}\\
LHCb (2015)    & --- & $0.336 \pm 0.027 \pm 0.030 $ \cite{Aaij:2015yra}\\
LHCb (2017)    & --- & $0.285 \pm 0.019 \pm 0.029$ \cite{wormser}\\
Average  & $0.407\pm 0.039\pm 0.024$ & $0.304\pm 0.013\pm 0.007$ \cite{hfag}\\
\hline
\hline
\end{tabular}
\caption{The SM predictions for and the data on $R(D)$ and $R(D^*)$.
  While \Babar~considers both charged and neutral $B$ decay channels,
  LHCb and Belle results, as quoted here, are based only on the
  analysis of neutral $B$ modes. For Belle, `Full' implies the inclusion 
  of both semileptonically and hadronically tagged samples, the last-mentioned 
  excluded otherwise. The average is from HFLAV \cite{hfag}. 
  }
\label{tab:RDRDst}
\end{center}
\end{table}

$\rd$ and $\rdst$ exceed their respective SM predictions by $2.3\sigma$ and $3.4\sigma$. 
Taking the correlation into effect, the discrepancy is at the level of $4.1\sigma$ \cite{hfag}. Thus,

\beq
\label{data:RD}
\rd = (1.36\pm 0.15\,)\times \rd_\text{SM},\qquad \qquad \rdst = (1.21
\pm 0.06\,)\times \rdst_\text{SM} \ .  \eeq At this point, let us
mention that recently two groups have published their own calculations
of $\rdst_{\rm SM}$, namely, $0.260\pm 0.008$~\cite{gambino} and
$0.257\pm 0.005$~\cite{sneha}.  While these estimates reduce the
tension slightly, to $2.3\sigma$~\cite{gambino} and
$2.9\sigma$~\cite{sneha} respectively, note that the error bars are
somewhat bigger. Given that the consequent changes in numerical
estimates are minimal, we will continue to use the results of
Ref.\ \cite{Kamenik:2008tj} in the main, but would also indicate the
change the results that would be wrought by the use of the new
calculations\footnote{It might be tempting to consider an average of
  the three calculations of $\rdst_{\rm SM}$. However, this cannot be
  effected in a straightforward manner as, on the one side, some of
  the theoretical errors are correlated, while, on the other some of
  the theoretical inputs are at slight variance. 
  On the other hand, Ref.\ \cite{deboer18} 
  shows that the soft electromagnetic corrections enhance the SM prediction for 
  $R(D)$ by 3-5\%. This reduces the tension with the SM slightly, but we have not taken
  this into account for our analysis, as a corresponding analysis for $R(D^*)$ is not yet 
  available.
  }.

\item 
Recently, the LHCb collaboration has observed the hint of another
discrepancy~\cite{lhc_b_future_paper} for the same
quark-level transition $b \to c \tau \bar{\nu}$ but in the $B_c$ meson
system. With only the spectator quark changing, the SM analysis would
be very similar to that for $R(D^*)$, with the main modification
accruing from the change in the phase-space factors. Much the same
would be true for a large class of new physics scenarios,
wherein the tensorial structure of the effective four-fermion operators remain 
essentially unchanged\footnote{Clearly, if
      the dominant new physics effect emanates from an operator
      different from that within the SM (as, for example, may happen
      in a theory with a light charged scalar), this would no longer be
      true.}.  Analyzing $3\, {\rm fb}^{-1}$ data, The LHCb Collaboration found
\beq
\label{data:RJpsi}
\rJpsi= 
\left\{\hskip-5pt
\barr{lcl}
\dsp 0.71  \pm 0.17\pm 0.18 &\hskip-5pt & ({\rm exp.}),
\\[2ex]
\dsp0.283 \pm 0.048 &\hskip-5pt & ({\rm SM})\, .
\earr
\right. 
\eeq
The SM prediction~\cite{Ivanov:2005fd,Dutta:2017xmj,Watanabe:2017mip}
includes the uncertainties coming from the $B_c\to J/\psi$ form
factors and, thus, is quite robust.  Given the relatively low
production cross section for the $B_c$ meson, the uncertainty in the
measurement is still large and the discrepancy is just below the
$2\sigma$ level. With the accumulation of more data, not only would
the statistical uncertainties reduce, even the systematics are
expected to come down on account of a better understanding of the
same. While the present level of the discrepancy is not a very
significant one, it is interesting to note that it points in the same
direction as the other charged-current decays.

\item
For the neutral current transitions (pertaining to $b \to s \ell^+
\ell^-$), we have \cite{1406.6482,Aaij:2017vbb}:
\beq
\label{data:RK}
\barr{rclcl}
\rk &=& \dis 0.745^{+0.090}_{-0.074}\pm 0.036 & \qquad &\dis q^2 \in [1:6]\, {\rm GeV}^2\,,\\[2ex]
\rkst^\text{\,low} &=&\dis {0.66}^{+0.11}_{-0.07} \pm 0.03 & \qquad &\dis  q^2 \in [0.045:1.1]\, {\rm GeV}^2\,,\\[2ex]
\rkst^\text{\,central} &=& \dis 0.69 ^{+0.11}_{-0.07} \pm 0.05  & \qquad &
        q^2 \in [1.1:6]\, {\rm GeV}^2\,.
\earr
\eeq
The SM predictions for both $\rk$ and $\rkst^\text{\,central}$ are
virtually indistinguishable from unity~\cite{sm-pred}, whereas for
$\rkst^\text{\,low}$ it is $\sim$ 0.9 (due to the finite lepton mass
effect). These calculations are very precise with
negligible uncertainties associated with them. Thus the measurements
of $\rk$, $\rkst^\text{\,low}$ and $\rkst^\text{\,central}$,
respectively, correspond to $2.6\sigma$, $2.1\sigma$ and $2.4\sigma$
deviations from the SM predictions.

\item
Another hint of deviation (at a level of more than $3\sigma$), 
for a particular neutral-current decay mode is evinced by
$B_s\to \phi\mu\mu$ ~\cite{Aaij:2015esa,Altmannshofer:2014rta,Straub:2015ica}.
\beq
\label{data:phimumu}
\frac{d~}{dq^2}{\rm BR}(B_s\to\phi\mu\mu){\Big|}_{q^2\in[1:6]\,{\rm GeV}^2}
= \left\{ 
    \barr{lcl}
    \dis \left(2.58^{+0.33}_{-0.31}\pm 0.08\pm 0.19\right) \times 10^{-8}~{\rm GeV}^{-2} & \quad & ({\rm exp.}) \\[2ex]
    \dis \left(4.81 \pm 0.56\right) \times 10^{-8}~{\rm GeV}^{-2} & & ({\rm SM})\,. 
  \earr
  \right.
\eeq
where $q^2 = m^2_{\mu\mu}$. Intriguingly, the $q^2$ region where this
measurement has relatively low error (and data is quoted) is virtually
the same as that for $\rk$ and $\rkst^\text{\,central}$.  This
measurement, thus, suggests strongly that the discrepancies in $\rk$
and $\rkst$ have been caused by a depletion of the $b \to s \mu^+
\mu^-$ channel, rather than an enhancement in $b \to s e^+ e^-$.  This
is further vindicated by the long-standing $P'_5$
anomaly~\cite{LHCb:2015dla} in the angular distribution of $B\to
K^*\mu\mu$, which again occurs in the central $q^2$ region, with the
mismatch between data and SM prediction being more than $3\sigma$.

\item
Such a conclusion, though, has to be tempered with the data for the
corresponding two-body decay, {\em viz.} $B_s\to \mu^+ \mu^-$.  The theory
predictions are quite robust now, taking into account possible
corrections from large $\Delta\Gamma_s$, as well as next-to-leading order (NLO) electroweak
and next-to-next-to-leading order (NNLO) QCD corrections. The small uncertainty essentially arises
from the corresponding Cabibbo-Kobayashi-Maskawa (CKM) matrix elements and the decay constant of
$B_s$.  The recent measurement of LHCb at a significance of
$7.8\sigma$ \cite{Bsmumu,BsmumuSM} shows an excellent agreement
between the data and the SM prediction:
\beq
\label{data:Bsmumu}
\text{BR}(B_s \to \mu^+ \mu^-)=
\left\{
\barr{lcl}
\dis \left(3.0\pm 0.6^{+0.3}_{-0.2}\right)\times {10}^{-9} & \quad & ({\rm exp.}), 
   \\[2ex]
 \dis \left(3.65 \pm 0.23\right) \times 10^{-9} & & ({\rm SM})\,,
\earr
\right.
\eeq
and hence puts very strong constraints on NP models, in particular on
those incorporating (pseudo-)scalar currents (this has also been 
noted in Ref.~\cite{Fleischer:2014jaa}). Of course, as the data
shows, a small depletion $\sim 20\%$ in the rates is still allowed,
and, indeed, to be welcomed.  NP operators involving only vector
currents do not affect this channel, but scalar, pseudoscalar, and
axial vector operators do. 

\end{itemize}

For future reference, we list here three further rare decay modes. 

\begin{itemize}
\item
Within the SM, the decay modes $b \to s \nu_i \bar\nu_i$ are
  naturally suppressed, owing to the fact that these are generated by
  either an off-shell $Z$-mediated diagram\footnote{The charged
      lepton modes, in contrast, are primarily mediated by off-shell
      photons.}  coupling to a flavour-changing quark-vertex, or
  through box-diagrams. Interestingly, the current upper bounds (summed
  over all three neutrinos) as obtained by the Belle collaboration \cite{belle17}, {\em viz.}
\beq
\label{data:MET}
{\rm BR}(B\to K\nu\bar\nu) < 1.6\times 10^{-5}\,,\qquad \quad
{\rm BR}(B\to K^*\nu\bar\nu) < 2.7\times 10^{-5}
\eeq
(both at 90\% C.L) are not much weaker than what is expected within the SM, namely~\cite{MET_SM}
\beq
\label{SM:MET}
\barr{rcl}
\dis {\rm BR}(B^+ \to K^+ \nu \bar\nu)_{\rm SM} & = & \dis 
(3.98 \pm 0.43 \pm 0.19) 
\times 10^{-6} \ , 
\\[2ex]
\dis {\rm BR}(B^+ \to K^{*0} \nu \bar\nu)_{\rm SM} & = & \dis 
(9.19 \pm 0.86 \pm 0.50) \times 10^{-6} \ .
\earr
\eeq
In other words, we have
\[
 \frac{{ \rm  BR}^{ \rm exp}_{K}}{{ \rm  BR}^{ \rm SM}_{K}} < 3.9\,,
\qquad \quad
 \frac{{ \rm  BR}^{ \rm exp}_{K^*}}{{ \rm  BR}^{ \rm SM}_{K}} < 2.7 \,.
\]

\item

The purely leptonic mode $B_s\to\tau\tau$ is yet to be observed but
LHCb put a 95\% confidence level bound on the BR \cite{Aaij:2017xqt}:
\beq
{\rm BR}(B_s\to\tau^+ \tau^-) < 6.8\times 10^{-3}\,.
\eeq
The SM prediction, of course, is way too small, $(7.73\pm 0.49)\times
10^{-7}$ \cite{BsmumuSM}.

\item
The limits on the rare lepton flavour violating
modes $B\to K\mu\tau$ \cite{pdg} are
\beq
\label{data:mutau}
{\rm BR}(B^+\to K^+\mu^+\tau^-) < 4.5\times 10^{-5}\,,\ \ 
{\rm BR}(B^+\to K^+\mu^-\tau^+) < 2.8\times 10^{-5}\,.
\eeq

\end{itemize}

 Finally, the direct production of 
$\tau^+\tau^-$ pairs at the LHC was well-measured
for both the 8 TeV and the 13 TeV runs~\cite{atlas-ditau}. This serves 
to constrain the WCs of any effective theory giving rise to a
$\tau^+\tau^-$ final state~Refs.\ \cite{1609.07138,1706.07808}.



\section{Operators relevant to the observables}
\label{sec:SM}

Having delineated, in the preceding section, the observables of
interest, we now proceed to the identification of the operators,
within the SM and beyond, responsible for effecting the transitions.
As the scale of NP surely is  above the
electroweak scale, we will talk in terms of effective current-current
operators that, presumably, are obtained by integrating out the heavy degrees of freedom, and not
confine ourselves within any particular model. To be precise, the
genesis of these operators will be left to the model builders.

Given the fact that, even within the SM, the neutral-current
decays under consideration occur only as loop effects, 
several current-current operators
would, in general, be expected to contribute to a given
four-fermion amplitude. However, as we shall soon see, certain
structures have a special role. To this end, we introduce a shorthand
notation: 
\beq
(x,y) \equiv \bar x_L \g^\mu y_L \ \ \ \  \forall\ \  x,y\,.
\eeq


\subsection{The \boldmath$ b\to c \tau \bar{\nu}_\tau$ transition}

This proceeds through a tree-level $W$-exchange in the SM. If the NP
adds coherently to the SM, one can write the effective Hamiltonian as
\beq
\label{eq:Hcnutau}
{\cal H}^\text{eff} = \frac{4G_F}{\sqrt{2}} V_{cb} \left(1+\Cn \right) \, 
   \left[(c,b) (\tau,\nu_\tau)\right]\,,
\eeq
where the NP contribution, parametrized by $\Cn$, vanishes in the SM
limit.  Using Eq.~\eqref{eq:RD}, we can write
\beq
\frac{{\rdrdst}_\text{SM+NP}}{{\rdrdst}_\text{SM}}= \left\vert 1+ \Cn\right\vert^2.
\eeq
Thus, to explain the data, one needs either small positive values, or
large negative values, of $\Cn$.  

On the other hand, if the NP involves an operator with a different Lorentz structure, or with 
different field content (like $b\to c\tau\bar\nu_\mu$), the addition would be incoherent in nature, 
and the relative phase of $\Cn$ is immaterial. More crucially, though,
this would alter the functional form of the differential width. An identical situation arises for $\rJpsi$.

%

\subsection{The \boldmath$ b\to s \mu^+ \mu^-$ transition}
\label{sec:btos}

Responsible for the FCNC decays $B\to K^{(*)} \mu^+ \mu^-$ and $B_s\to
\phi \mu^+ \mu^-$, within the SM, this transition proceeds, primarily,
through two sets of diagrams, viz. the ``penguin-like'' one (driven
essentially by the top quark) and the ``box'' (once again, dominated
by the top).  It is, thus, convenient to parametrize the ensuing
effective Hamiltonian as
\bea
{\cal H}^\text{eff} = \frac{-4G_F}{\sqrt{2}} \, V_{tb} \, V_{ts}^*\, 
              \sum_{i}C_i(\mu) \mathcal{O}_i(\mu)\,,
\eea
where the relevant operators are 
\beq
\label{eq:opbtos}
\mathcal{O}_7= \frac{e}{16\pi^2}m_b \left(\bar{s} \sigma_{\mu\nu} P_R b\right) 
   F^{\mu \nu} \,,\ \ 
\mathcal{O}_9= \frac{e^2}{16\pi^2} \left(\bar{s} \g_\mu P_L b\right) \left(\bar{\mu} \g^\mu \mu\right) \,,\ \ 
\mathcal{O}_{10}= \frac{e^2}{16 \pi^2} \left(\bar{s} \g_\mu P_L b\right) \left(\bar{\mu} \g^\mu\g_5 \mu\right)\,.
\eeq
The WCs, matched with the full theory at $m_W$ and then run down to
$m_b$ with the renormalisation group equations at the
next-to-next-to-leading logarithmic (NNLL) accuracy
\cite{Altmannshofer:2008dz}, are given in the SM as $C_7 = -0.304
\,,\ \ C_9 = 4.211 \,,\ \ C_{10}= -4.103 \,.$ If the NP operators are
made only with (axial)vector currents, one can denote the modified WCs
as 
\beq 
C_9 \to C_9+ \Cnine= 4.211 + \Cnine\,,\ \ \ 
C_{10} \to C_{10}+ \Cten= -4.103 + \Cten\,.  
\eeq 
The consequent normalized differential branching fraction for the $B\to K
\mu^+ \mu^-$ decay in terms of the dimuon invariant mass squared,
$q^2$, is given by \bea \frac{1}{\Gamma_0} \frac{d\Gamma(B\to K \mu^+
  \mu^-)}{dq^2}\! &=&\! 2 \lambda^{1/2}(m_B^2,m_K^2,q^2)\sqrt{1-
  \frac{4 m_\mu^2}{q^2}} \Bigg\{ \frac{1}{6}\lambda(m_B^2,m_K^2,q^2)
\left(1+ \frac{2 m_\mu^2}{q^2}\right) \left( |F_A|^2+ |F_V|^2 \right)
\nonumber \\ \!&+& \! q^2 |F_P|^2 + 4 m_\mu^2 m_B^2 |F_A|^2 +2 m_\mu
\left(m_B^2- m_K^2 +q^2\right) \text{Re}(F_PF_A^*) \Bigg\}, \eea where
$$
\Gamma_0= \frac{G_F^2 \alpha^2}{ 2^9 \pi^5 m_B^3}|V_{tb} V_{ts}^*|^2,~{\rm and}~~
\lambda(a, b,c)\equiv a^2 + b^2 + c^2 -2 (a b + b c + a c)
$$
is the famed phase-space factor.
The functions $F_i$ depend on the WCs and $q^2$
dependent form factors $f_{0,+,T}$ of the $B\to K$
transition \cite{Ball:2004ye}, namely
\beq
\barr{rcl}
F_P &=& \dis -m_\mu \left(C_{10}+ \Cten\right) \,
        \left[f_+(q^2) - \frac{m_B^2-m_K^2}{q^2}\left(f_0(q^2)-f_+(q^2)\right) 
          \right] \ ,   
\\[3ex]
F_A &=& \dis \left(C_{10}+ \Cten\right) \, f_+(q^2) \ ,
\\[2.5ex]
F_V & = & \dis
\left(C_9+ \Cnine  \right)f_+(q^2) + 
         2C_7 \, m_b \, \frac{f_T(q^2)}{m_B+m_K} \ .
\earr
\eeq
The differential distribution for the $B \to V \ell \ell $
mode, where $V$ stands for a $K^*$ or $\phi$ meson, can be expressed
in terms of certain angular coefficients $I_i$~\cite{Kruger:1999xa} as
\bea
\frac{d\Gamma(B\to V \mu^+ \mu^-)}{dq^2} = \frac{1}{4} \left[3 I_1^c(q^2) + 6 I_1^s(q^2)-I_2^c(q^2)-2I_2^s(q^2)  \right]
\eea
The coefficients $I_i$ are functions of the transversity
amplitudes $\mathcal{A}^{L,R}_{\lambda,t}$ where $\lambda$ denotes the
three states of polarisations of the meson $V$, and $L$ and $R$ denote
the left and right chirality of the lepton current, respectively. We
refer to Appendix~\ref{sec:appendix} for the detailed expressions of
$I_i$ and $\mathcal{A}^{L,R}_{\lambda,t}$.

\subsection{The \boldmath$ b\to s \nu \bar{\nu}$ transition}
\label{sec:MET}
Quite akin to the preceding case, this
transition (which governs the $B\to K^{(*)} \nu \bar{\nu}$ decay) 
proceeds through both $Z$-penguin and box diagrams. 
Unless NP introduces right-handed neutrino fields, the 
low energy effective Hamiltonian may be parametrised
by~\cite{MET_SM}
\bea
{\cal H}^\text{eff} = -\frac{4G_F}{\sqrt{2}} \, V_{tb} \, V_{ts}^*\,
   \frac{\alpha_{\rm em}}{4\pi} \, C_L^\text{SM} \, \left(1+ \Cnu\right)\, 
\times 2 \, (s,b)(\nu,\nu),
\eea
where $\alpha_{\rm em}$ is the fine structure constant and \Cnu
denotes the NP contribution. Including NLO QCD corrections and the two
loop electroweak contribution, the SM WC is given by
$C_L^\text{SM}=-X_t/s_w^2$ where $X_t= 1.469 \pm
0.017$~\cite{MET_SM_old,MET_SM}.

\subsection{The two-body decay rates}

The branching fraction for $B_s^0 \to \ell^+ \ell^-$, where $\ell$ is any charged lepton,
 can be written, at the leading order, as
\bea
{\rm BR}(B_s^0 \to \ell^+ \ell^-)= \frac{G_F^2 \alpha^2 m_{B_s} \tau_{B_s} f_{B_s}^2 m_\ell^2}
{16 \pi^3} |V_{tb} V_{ts}^*|^2 \sqrt{1- \frac{4 m_\ell^2}{m_{B_s}^2}}~ \left |C_{10} + \Cten \right|^2\,,
    \label{bs_mumu_expr}
\eea
while that for $B_c^-\to \tau^- \nu_\tau$ is given by
\bea
{\rm BR}(B_c^-\to \tau^- \nu_\tau)= \frac{G_F^2 m_{B_c} \tau_{B_c^-} f_{B_c}^2  m_\tau^2 }{8\pi} 
|V_{cb}|^2\left(1- \frac{m_\tau^2}{m^2_{B_c}} \right)^2 \left|1+ \Cn\right|^2\,,
\eea
where $m_{B}$, $\tau_{B}$ and $f_{B}$ are the mass, lifetime and
decay constant of the relevant $B$ meson respectively.
We assume an identical operator structure leading to coherent 
addition of the amplitudes.



\section{Semi-realistic scenarios}
\label{sec:model}

In the spirit of effective theories, a ``model'' in our discussions
would correspond to only a combination of (at most) two
four-fermi operators at the scale $m_b$. We, expressly, do not venture to obtain the
ultraviolet (UV) completion thereof, leaving this for future studies. We also assume that 
the NP scale is low enough, maybe at a few TeV, so that higher-loop corrections do not
generate any new operator of significant strength, leading to, {\em e.g.}, the purely leptonic 
decay $\tau\to 3\mu$. Our
aim, thus, is to identify the smallest set of couplings that are
phenomenologically viable. To begin with, we present a model 
that we call semi-realistic, because while it can explain all other anomalies, it is inconsistent 
with the constraint coming from the decay $B_s\to\tau^+\tau^-$.

This instructive exercise would allow us to pinpoint the structure that NP must lead to so as 
both explain the anomalies as well as satisfy all other constraints. 
We will follow this, in the next section, by constructing appropriate 
realistic scenarios. 

Clearly, any such operator should, at the very least, respect the
(gauged) symmetries of the SM. On the other hand, since the data
explicitly calls for lepton-flavour violation, the latter cannot be a
symmetry of the NP operator(s). Such a violation can accrue from a
plethora of NP scenarios, such as models of (gauged) flavour,
leptoquarks (or, within the supersymmetric paradigm, a breaking of
$R$-parity) etc. Let us here  investigate a structure that
could, in principle, be motivated from such theories.

The data in question calls for the effect of NP to be concentrated in
effective operators straddling the second and third generations. On
the other hand, any such effective Hamiltonian accruing from a NP
scale higher than the electroweak scale (as it must be, on account of
the colliders, past and present, not having seen such resonances), can
only be written in terms of the weak-interaction eigenstates~\cite{ggl}. The
breaking of the electroweak symmetry, aided by non-diagonal Yukawa
couplings, would, in general, induce extra operators even if we
started with a single one. While some of these effects would be
trivial (and aligned with the usual CKM
rotations), this is not necessarily true for all.  We will exploit
this in striving to explain all the anomalies starting from a minimal
set.  In particular, our operators will involve second and third
generation quark fields, to account for the $b\to c$ and $b\to s$
transitions, but only the third generation lepton fields, with the
charged lepton state appropriately rotated to give rise to the mass
eigenstates of $\mu$ and $\tau$ leptons.

\subsection{The could-have-been model}
   \label{subsec:1}

Let us consider the following set of operators: 
\beq
\label{eq:op3}
\hspace*{-0.8cm }{\rm Model~ I:}\qquad  \mathcal{O}_{\rm I} = \sqrt{3} \, A_1 \, (\bar Q_{2L} \g^\mu L_{3L})_3 \, 
(\bar L_{3L} \g_\mu Q_{3L})_3 -
2 \, A_2 \, (\bar Q_{2L} \g^\mu L_{3L})_1 \, (\bar L_{3L} \g_\mu Q_{3L})_1 \,.
\eeq
where $A_{1,2}$ are the\footnote{Here, as well as later, the
  coefficients $A_i$, of dimension (mass)$^{-2}$, are considered to be
  real. This simplifying choice eliminates any source of CP-violation
  from beyond the SM.}  WCs, and $Q_i$ and $L_i$ denote the usual
$i$-th generation (and weak-eigenstate) $SU(2)_L$ quark and lepton
doublet fields respectively. The subscript `3' and `1' denote that the
currents are $SU(2)_L$ triplets and singlets, respectively, whereas
the factor of $\sqrt{3}$ has been introduced explicitly to account for
the Clebsch-Gordan coefficients.  It should be noted that only the second and
third generation quark doublets and the third generation 
lepton doublet alone are involved in Eq.~\eqref{eq:op3}, as
mentioned earlier. Considering the simplest of field rotations for the
left-handed leptons from the unprimed (flavour) to the primed (mass)
basis, namely
\beq
\label{eq:rot}
\tau = \ct \, \tau' + \st \, \mu' \ ,
\qquad \quad 
\nu_\tau = \ct \, \nu_\tau' + \st \, \nu_\mu'\,.
\eeq
terms with the potential to explain the $b\to s \mu \mu$ anomalies are generated.

%
%
The best fit values for these can be obtained by effecting a $\chi^2$-test defined
through
\bea
\label{eq:chisquare}
\chi^2= \sum_{i=1}^{8} \frac{\left(\mathcal{O}_i^\text{exp}-\mathcal{O}_i^\text{th}
\right)^2}{\left(\Delta \mathcal{O}_i^\text{exp}\right)^2+ \left(\Delta \mathcal{O}_i^\text{th}\right)^2}
\eea
where $\mathcal{O}_i^\text{exp}$ and $\Delta \mathcal{O}_i^\text{exp}$
are the experimental mean and $1\sigma$ uncertainty in the
measurements and $\Delta\mathcal{O}_i^\text{th}$ is the theoretical
uncertainty in the observables. 
The observables $\mathcal{O}_i^\text{th}$ are calculated within Model I and thus
depend on the model parameters. We include a total of eight
measurements for the evaluation of $\chi^2$, namely, $\rd,~ \rdst$
(from Eq.~\eqref{data:RD}), $\rJpsi$ (from Eq.~\eqref{data:RJpsi})
$\rk,~\rkst^\text{\,low},~\rkst^\text{\,central}$ (from
Eq.~\eqref{data:RK}), the integrated $\text{BR}(B_s\to\phi\mu\mu)$ for
the range $q^2\in[1:6]\,{\rm GeV}^2$ (from Eq.~\eqref{data:phimumu})
and $\text{BR}(B_s\to\mu\mu)$ (from Eq.~\eqref{data:Bsmumu}). 
The theoretical uncertainties $\Delta\mathcal{O}_i^\text{th}$, however small, 
are taken into account for all observables. For our numerical analysis, we use
\[
V_{cb}=0.0416 \ , \qquad V_{tb}V_{ts}^*=-0.0409 \ , \qquad s_w^2=0.22\,,
\]
and find, for the SM, $\chi^2_{\rm SM}/{\rm d.\,o.\,f.} \simeq 6.1$.
For Model I, on the other hand, the minimum value is $\chi^2_{\rm min}/{\rm d.\,o.\,f.} \simeq 1.5$
denoting a remarkable improvement from the SM value. The fit results are
\beq
A_1 = 0.028~{\rm TeV}^{-2}\,, \ \ A_2 = -2.90~{\rm TeV}^{-2}\,, \ \ \vert\sin\theta\vert = 0.018\,,
\eeq 
leading to
\beq
{\rm BR}(B^+ \to K^+\nu\bar\nu) = 6.1\times 10^{-6}\,,\ \ 
{\rm BR}(B \to K^*\nu\bar\nu) = 1.4\times 10^{-5}\,.
\eeq
Note that the small value of $A_1$ means that at the matching scale of the effective theory,
$A_1$ could very well be zero, making this a two-parameter model ($A_2$ and $\st$).
The fit would have improved significantly if we exclude $\rkst^\text{\,low}$ from our analysis. 
The BRs for $B\to K\mu\tau$ and $B\to K^*\mu\tau$~\cite{Kmutau} 
increase to $2.1\times {10}^{-5}$ and $3.6\times {10}^{-5}$ respectively, just
below the current observed limit (see Eq.~\eqref{data:mutau}).

The NP contribution to the WCs \Cnine, \Cten and \Cn come out to be
\beq 
\Cnine = -\Cten = -0.61 \ , \qquad \Cn = -2.11 \ .  
\eeq 
Not only is this completely consistent with the global fit results for
$b\to s$ transitions, but also provides an explanation for the well
known $P_5^\prime$~\cite{LHCb:2015dla} anomaly as it calls for an
(axial)vector contribution to the muon mode with similar coupling
strength~\cite{Altmannshofer:2014rta}.

We come, finally, to the decay mode that rules out this model,  namely $B_s \to \tau^+ \tau^-$. The
theoretical expression for this mode is given in
Eq.\ (\ref{bs_mumu_expr}). It can be ascertained that the typical
value of the BR as predicted within Model I is
{\em an order of magnitude} higher than the LHCb limit quoted
before~\cite{Aaij:2017xqt}.  Indeed, the only way the $B_s \to \tau^+ \tau^-$
constraint can be satisfied is to tweak the values of the WCs to an
extent that the best fit value for $R_{K^*}$ is nearly $2\sigma$ away
from the global average, thereby negating the very essence of the
effort.  Depending on the structure of the UV-complete
theory, one may even have a similar large
contribution, in stark contrast to the data, to the mass difference
$\Delta M_s$ for the $B_s$ system. This issue is discussed at a later stage.

At the same time, let us note that the LHCb collaboration has, very
recently, announced measurement of $R(D^*)$ through the 3-prong decay
of the $\tau$~\cite{LHCb_rdst_3prong}, 
with this particular data being
consistent with the SM expectations at about $1\sigma$. More
importantly, while the global average of $R(D^*)$ reduces, its
deviation from the SM value actually {\em increases} marginally to
$\sim 3.4\sigma$~\cite{LHCb_rdst_3prong} on account of the improved
precision\footnote{We do not include this measurement in our analysis
  as the corresponding error correlations with $\rd$ have not yet been
  worked out.}. This movement of the global average allows for the
adoption of smaller values for $A_i$, thereby reducing the tension
with $B_s \to \tau^+ \tau^-$.  Indeed, if $R(D^*)$ were to be entirely
dominated by this new measurement, or the BR for $B_s\to \tau^+\tau^-$
were found to be larger than the LHCb limit, the tension would be
reduced to acceptable levels and this model would have worked fine.

Before we proceed further, let us mention that one can construct similar models, like
\bea
\label{eq:op1}
\hspace*{-1cm }{\rm Model~ II:}\qquad \mathcal{O}_{\rm II} &=& 
-\sqrt{3} \, A_1 \, (\bar Q_{2L} \g^\mu Q_{3L})_3 \, 
               (\bar L_{3L} \g_\mu L_{3L})_3 
  + \sqrt{3} \, A_2 \, (\bar Q_{2L} \g^\mu L_{3L})_3 \, 
                  (\bar L_{3L} \g_\mu Q_{3L})_3 \ ,
\\[2ex]
\label{eq:op2}
\hspace*{-1cm }{\rm Model~ III:}\qquad  \mathcal{O}_{\rm III} &=& 
-\sqrt{3} \, A_1 \, (\bar Q_{2L} \g^\mu Q_{3L})_3 \,
(\bar L_{3L} \g_\mu L_{3L})_3 +
2 \, A_3 \, (\bar Q_{2L} \g^\mu Q_{3L})_1 \, (\bar L_{3L} \g_\mu L_{3L})_1 \, ,
\eea
while all these models satisfy every constraint and yield similar $\chi^2_{\rm min}$,
they all fail to keep BR($B_s\to\tau^+\tau^-$) within the allowed limit.

 We, nonetheless, endeavour below to identify a scenario that is not
 dependent on wishful thinking, as aforementioned.



\section{Realistic models}
\label{sec:model_new}

The strong constraints from $\Delta M_s$ and $B_s \to \tau^+ \tau^-$,
as seen in the preceding section, emanated from the fact that, in each
of the cases, we generated a substantial $C_{10}^{\rm NP}$, namely a
NP contribution to the operator $\mathcal{O}_{10}$ for the
tauonic mode (in Eq.~\eqref{eq:opbtos})
\[
    \frac{e^2}{16 \pi^2} \left(\bar{s} \g_\mu P_L b\right) \left(\bar{\tau} \g^\mu\g_5 \tau\right) \ .
\]
Indeed, we consistently had $C_{10}^{\rm NP} = - C_9^{\rm NP}$. While
the operator $\mathcal{O}_9$ has only a vanishing contribution to the decay,
the situation is very different for $\mathcal{O}_{10}$. Although the latter
contribution suffers a chirality suppression, it is still substantial
for $B_s \to \tau^+ \tau^-$. Thus, consistency with this mode
would require $C_{10}^{\rm NP}$ to be small.

On the other hand, a substantial change in $R_{K^{(*)}}$ would require 
at least one of  $C_{10}^{\rm NP}$ and $C_9^{\rm NP}$ to be substantial. 
In other words, we must break the relation $C_{10}^{\rm NP} = - C_9^{\rm NP}$, 
which was a consequence of working, thus far, with left-handed currents
alone. Learning from the lessons of the preceding section, we now consider 
simple variants of the models already introduced. 

\subsection{Model IV}
    \label{subsec:Ia}
Consider the following set of operators
\beq
\barr{rcl}
{\cal O}_{\rm IV} & = & \dis 
   \sqrt{3} \, A_1 \, \left[
- (\bar Q_{2L} \g^\mu Q_{3L})_3 \, (\bar L_{3L} \g^\mu L_{3L})_3
+ \frac{1}{2} \, (\bar Q_{2L} \g^\mu L_{3L})_3 \, (\bar L_{3L} \g^\mu Q_{3L})_3
     \right]
\\[2ex]
&+ & \dis
\sqrt{2} \, A_5 \,  (\bar Q_{2L} \g^\mu Q_{3L})_1 \, (\bar \tau_R \g^\mu \tau_R) \ ,
\earr
    \label{eq:Ia}
\eeq
where the new WC $A_5$ (note that the factor of $\sqrt{2}$ 
is a Clebsch-Gordan factor) parametrizes the strength of the right-handed 
tauonic current. In terms of the component field, this reduces to 
\beq
\barr{rcl}
{\cal O}_{\rm IV} 
 & = & \dis \frac{3 \, A_1}{4} \, (c, b) \, (\tau, \nu_\tau)
+ \frac{3 \, A_1 }{4} (s, b) (\tau, \tau) + A_5 \, (s, b) \, \{\tau, \tau\} 
\\[2ex]
& + & \dis \frac{3 \, A_1}{4} \, (s, t) \, (\nu_{\tau}, \tau) 
   + A_5 (c, t) \{\tau, \tau\}
   + \frac{3 \, A_1 }{4}(c, t) \, (\nu_\tau, \nu_\tau)
\earr
    \label{eq:Ia_comp}
\eeq
where we have introduced the shorthand notation 
\beq
\{x,y\} \equiv \bar x_R \g^\mu y_R \ \ \ \  \forall\ \  x,y\, , 
\eeq
and the terms in the second line of eqn.(\ref{eq:Ia_comp}) are
irrelevant for the processes in hand.  Clearly, $A_5 \simeq 3 A_1 / 4$
would suppress any NP contribution to $B_s \to \tau^+
\tau^-$. Simultaneously, this will
automatically generate a tiny contribution to $B\to
K^{(*)}\mu^+\mu^-$, comparable with the SM contribution, without
needing to tune the leptonic mixing angle to an unnaturally small value. Note that we
have already imposed the symmetry that had led to the suppression of
the $B \to K^{(*)} \bar\nu \nu$ modes.

While the introdution of the right-handed current opens the
possibility of introducing an independent mixing matrix for the
right-handed leptons, we eschew this in the interest of having the
fewest parameters in the NP sector. This would also be the case for
the other model discussed below.

\subsection{Model V}
    \label{subsec:IIa}
The analogous change in Model II and in Model III would, interestingly,
result in the same set of operators as
\beq
\barr{rcl}
{\cal O}_{\rm V} & = & \dis 
   -\sqrt{3} \, A_1 \, (\bar Q_{2L} \g^\mu Q_{3L})_3 \, 
               (\bar L_{3L} \g^\mu L_{3L})_3 
    +  A_1 \, (\bar Q_{2L} \g^\mu Q_{3L})_1 \, 
                 (\bar L_{3L} \g^\mu L_{3L})_1 
\\[2ex]
& + & \dis 
\sqrt{2} \, A_5 \, (\bar Q_{2L} \g^\mu Q_{3L})_1 \, (\bar \tau_R \g^\mu \tau_R)
\earr
    \label{eq:IIa}
\eeq
leading to 
\beq
\barr{rcl}
{\cal O}_{\rm V} 
& = & \dis 
A_1 \, (c, b) \, (\tau, \nu_\tau) 
+ A_1 \, (s, b) \, (\tau, \tau)
+ A_5\,  (s, b) \, \{\tau, \tau\} 
\\[2ex]
&+ & \dis 
A_1 \, (s, t) \, (\nu_{\tau}, \tau) 
+ A_1 \,  (c, t) (\nu_\tau, \nu_\tau)
+ A_5 \, (c, t) \, \{\tau, \tau\}
\earr
    \label{eq:IIa_comp}
\eeq
with the first line containing all terms of relevance. Once again, the (symmetry) relation between the coefficients of the first two
operators helps evade constraints from $B \to K^{(*)} \bar\nu
\nu$. Moreover, the very structure of Eq.\ (\ref{eq:IIa_comp})
suggests that $A_5 \simeq A_1$ would be preferred by $B_s \to \tau^+
\tau^-$.


\section{Results}
\label{sec:results}

The fitting of NP operators progress in exact analogy with that in
Sec.\ \ref{subsec:1}, with the relaxation of the condition
$C_{10}^{\rm NP} = - C_9^{\rm NP}$. Whereas the $\chi^2$ (defined in
Eq.~\eqref{eq:chisquare}) is still minimized for $|\sin\theta| \simeq
0.018$, such a solution would not simultaneously satisfy both of
BR$(B^+ \to K^+ \mu^- \tau^+)$ as well as BR$(B_s \to \tau^+ \tau^-)$
and yet lead to a satisfactory solution for the other variables.
Consequently, the best fit is obtained for a slightly different value,
namely\footnote{This is equivalent to imposing a penalty function and
  effecting a constrained minimisation of the $\chi^2$.}
\beq
| \sin\theta | \simeq 0.016
\eeq
applicable to each case. We now have $\chi^2_\text{min}/{\rm d.\,o.\,f.}= 1.7$, a rise 
of 1.3 from the unconstrained minimum. The new best fit values of the 
WCs are 
\[ 
C_9^\text{NP}=-0.35 \ , \qquad 
C_{10}^\text{NP}=0.55 \ , \qquad 
C^\text{NP}=-2.11 \ .
\]
This shows a marked improvement\footnote{Had we chosen to work with
  $R(D^*)_{\rm SM}$ estimates of refs.\cite{gambino, sneha} instead,
  the SM value for the $\chi^2$ would have been $43.3$ and $45.7$
  respectively. On the inclusion of the new operators, the values
  would have been 9.4 and 9.1 respectively.}  from the SM value of
$\chi^2/{\rm d.\,o.\,f.} = 6.1$. It should be noted that even this low
value of $\chi^2_{\rm min}/{\rm d.\,o.\,f.}= 1.7 $ is dominated by a
single measurement, namely, $\rkst^\text{\,low}$. Indeed, for the best
fit points of the NP parameter space, we have
$\rkst^\text{\,low}=0.83$ which is a little more than $1\sigma$ way
from the LHCb measurement.  This is quite akin to the various other
studies in the literature~\cite{rknew} who too have concluded that an
agreement to this experimental value to better than $1 \sigma$ is not
possible if the NP contribution can be expressed just as a
modification of the SM Wilson coefficients. Rather, it necessitates
the introduction of a new dynamical scale altogether.  Having
determined $\sin\theta$ as above, we choose, for illustrative
purposes, to relax the conditions imposed by ${\rm BR}(B^+ \to K^+
\mu^- \tau^+)$ as well as ${\rm BR}(B_s \to \tau^+ \tau^-)$.  This
would allow us to choose the best fit values for the parameters in
each of the two cases. These are displayed in
Table~\ref{Table:fit_values}.

\begin{table}[h]
\centering
\begin{tabular}{ |c| c |c |  }
	\hline \hline
	 Best fit points& Model IV &  \rule{0pt}{4ex}Model V 
\\ \hline
$|$sin$\theta |$ & \rule{0pt}{4ex} $0.016$ & $0.016$ 
\\
$A_1$ in TeV$^{-2}$ & $ -3.88$   & $ -2.91$ 
\rule{0pt}{4ex} \\
$A_5$ in TeV$^{-2}$& $ -2.61$   & $0.66$ 
\rule{0pt}{4ex}  \\ \hline
\hline 
\end{tabular}
\caption{Central values of the fitted parameters for the two models. }  
   \label{Table:fit_values}
\end{table}

\begin{figure}[!ht]
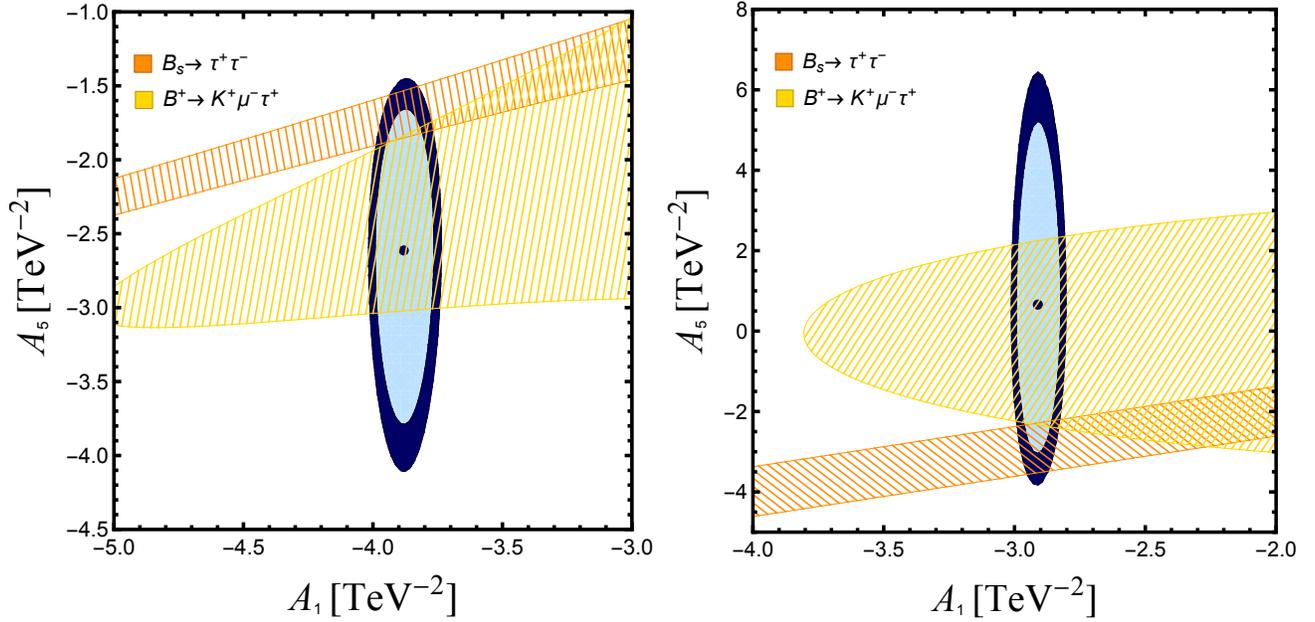

	\begin{center}
\includegraphics[width=0.502\linewidth]{mod1a.pdf}
\includegraphics[width=0.49\linewidth]{mod2a.pdf}
\caption{The viable parameter space for two different models i.e.,
  model IV and model V are shown in left and right panel, respectively. The point represents the minimum 
  of the $\chi^2$, whereas the
  light and dark blue ellipses denote $95\%$ and $99\%$ C.L. 
  regions, respectively. The orange and yellow shaded regions are
  allowed by the BR$(B_s \to \tau^+ \tau^-)$ and BR$(B^+
  \to K^+ \mu^- \tau^+)$ bounds. The overlaps denote the 
  finally allowed portion of the parameter space. }
	\label{fig:final_fit}
	\end{center}
\end{figure}

In Fig.\ \ref{fig:final_fit}, we depict the $95\%$ and $99\%$
C.L. bands, in the plane of the WCs, around the new best fit point.  Also shown are the regions in the
parameter space allowed by the upper limits on ${\rm BR}(B_s \to \tau^+
\tau^-)$ (the orange shaded region) and ${\rm BR}(B^+ \to K^+ \mu^- \tau^+)$
(the yellow shaded region). The former, as expected, is truly restrictive.
The overlap region, thus, denotes the viable parameter space at a
given confidence level.

The observables at the overlap region, corresponding to a least value
of $\chi^2/{\rm d.\,o.\,f.} \simeq 2.6$, are given by
\beq
\barr{rcl}
\rk & = & 0.80 \ ,\\[1ex]
\rkst^{\text{low}} & = & 0.88 \ , \\[1ex]
\rkst^{\text{central}} & = & 0.83 \ , \\[1ex]
\rd & = & 0.36 \ , \\[1ex]
\rdst & = & 0.30 \ , \\[1ex]
\rJpsi & = & \dis 0.34 \ , \\[1ex]
\dsp\frac{d}{dq^2}\text{BR}(B_s\to\phi\mu\mu) & = & 3.8 \times 10^{-8}\,{\rm GeV}^{-2} ~~ \mbox{for} \quad q^2\in[1:6]\,{\rm GeV}^2 \ .
\earr
\eeq 
 From Fig.\ \ref{fig:final_fit}, one might
be led to think that the models IV and V are consistent only at 95\%
C.L. or worse, but this is deceptive. It should be noted that the
contours shown in Fig.~\ref{fig:final_fit} are not drawn around the
absolute minimum of $\chi^2$, which, in any case, is incompatible with
other data, namely, $B^+ \to K^+ \mu^- \tau^+ $ and $B_s \to \tau^+
\tau^-$. Yet, the $\chi^2$ values corresponding to the overlap region,
namely $\ltap 15\, (18)$ for the 95\% (99\%) C.L. bands, are much
better than that obtained within the SM, which is $\sim 49$. In other
words, the improvement is remarkable.

It has been shown in Ref.~\cite{Alonso:2016oyd} that the
inclusion of left-handed NP current in the $b\to c \tau
\bar{\nu}_\tau$ transition to explain the $\rdrdst$ anomalies, does
not jeopardize the lifetime of the $B_c$ meson significantly, although
it opens up an annihilation mode $B_c\to\tau\nu$ for its decay. Our
results for NP coefficients correspond to a
modification of $\sim 3 \%$ of the lifetime and is well within the
allowed limit.
  
We also make a couple of strong predictions which should be tested in
LHCb in near future.  First, for both the models under consideration,
the similar ratio in $B_s \to \phi \ell^+\ell^-$ mode, namely,
$R_\phi\equiv{\rm BR}(B_s\to \phi \mu^+\mu^-)/ {\rm BR}(B_s\to \phi
e^+ e^-)$ should be less than unity and is predicted to be $\simeq
0.83$ for the range $q^2\in[1:6]\,{\rm GeV}^2$.  Second, as the
allowed region almost saturates the bounds arising from the modes
${\rm BR}(B_s \to \tau^+ \tau^-)$ and ${\rm BR}(B^+ \to K^+ \mu^-
\tau^+)$, they should also be observable in near future, and so should
be $B\to K^*\tau\mu$.  Apart from this, several
anomalous top decay channels may be probed at the LHC or the next
generation $e^+e^-$ collider. Each of these
predictions provides independent modes to both
    test and falsify the scenarios proposed.

%
%

\subsection{Caveat: Quantum corrections and new operators}

While the issue of too large a ${\rm BR}(B_s\to \tau^+ \tau^-)$ faced by
the semi-realistic scenarios discussed in Sec.~\ref{sec:model} was
taken care of\footnote{Note that this could not have been solved by
    introducing another intermediate particle. If the said particle
    were light, it should have been discovered by now. If it were
    heavy instead, the corresponding field could have been integrated
    out leading to a modification in the Wilson coefficients in the
    effective Lagrangian. Thus, such a step could, at best, have
    mimicked the situations discussed in Sec.~\ref{sec:model_new}.} in
the models proposed in Sec.~\ref{sec:model_new}, a further issue
remains. 

The operators that we discussed can also, in principle, generate new
 contributions to $B_s$--$\bar{B_s}$ mixing. Consider, {\em e.g.} the $\tau$-loop diagram
 contributing to the effective $\Delta B = \Delta S = 2$ operator. The amplitude  is
 formally a quadratically divergent one. Thus, to calculate it one
 needs to introduce a cutoff $\Lambda$ which could be estimated by
 parametrizing $A_i = a_i / \Lambda^2$ with $a_i \sim {\cal O}(1)$. In
 other words, the fit dictates $\Lambda$ to be a few TeV at
 most. Using $\Lambda$ as a cutoff would, naively, generate a WC that
 only scales with $\Lambda^{-2}$. In other words, we seemingly
 have\cite{Choudhury:2012hn}
\beq
\frac{\zeta}{\Lambda^2} \, (\bar b \, \gamma_\mu \, P_L \, s) \, 
(\bar b \, \gamma^\mu \, P_L \, s) \qquad \quad 
\zeta \sim {\cal O}\left(\frac{a_i^2}{\pi^2}\right)\,.
    \label{bbss_op}
\eeq
Accepting this, the experimental measurement of the mass difference
$\Delta M_s$ imposes a rather strong constraint, namely, $A_1\, \ltap \,
0.1\, {\rm TeV}^{-2}$~\cite{Choudhury:2012hn}, and this limit falls
linearly with increasing $\Lambda$.  This apparent conflict with our
fit results can be evaded if there are more contributions to the
$B_s$--$\bar{B_s}$ mixing. A trivial example is provided by
postulating a ``tree-level'' operator with a form identical to that
above and a coefficient with a sign opposite to $\zeta$ above.  A more
interesting alternative could be to appeal to some
yet-to-be-discovered symmetry of the full theory that cancels out, at
least approximately, all the quantum corrections to the existing set
of dimension-6 operators that may lead to a sizable $\Delta
M_s$. Either solution could, of course, be termed a slightly
fine-tuned one.

Before we delve deeper into this problem, it behoves us to consider
the very calculation of $\zeta$ indicated above. One criticism is that
the result is dependent on the regularisation prescription and a
different one could have resulted in a markedly different $\zeta$. A
more subtle issue pertains to the very nature of such calculations in
an effective theory. Indeed, an effective Lagrangian is presumably the
result of either having integrated out the heavy fields in a more
fundamental theory or having incorporated (some of) the quantum
corrections to yield an effective action.  If the local Lagrangian
under consideration is to be thought of as the lowest order
approximation in an expansion, further quantum corrections due to the
UV physics alone can only result in corrections to the WCs
and not generate any new terms in the Lagrangian\footnote{This
  argument does not hold for nonlocal terms, or, equivalently, terms
  characterized by nonanalytical forms in the momentum space. These,
  however, are not of interest to us.}. Such new terms should arise
only when quantum corrections to low-energy physics are taken into
account.  

In the present context, if say the operator ${\cal O}_{V}$ were the
result of a $Z'$ exchange, then the operator of Eq.\ (\ref{bbss_op})
should have been generated at the same level as ${\cal
  O}_{V}$. Suppressing $B_s$--$\bar B_s$ mixing would, then, require
that the $bsZ'$ coupling be far smaller than the $\tau\tau Z'$ one.
Calculating the $\tau$-loop would, then, be unnecessary and largely
meaningless. On the other hand, imagine that the operator ${\cal
  O}_{V}$ was generated as a combined effect of a slew of coloured
fields (with the displayed form being the result of a final Fierz
rearrangement). In such a case, the operator of Eq.\ (\ref{bbss_op})
would be generated only when mixed loops involving both these coloured
(and heavy) fields as well as the SM bosons. Intricately woven with
this are dependence on the light masses and the analogue of the GIM
cancellations\footnote{To put this analogy into perspective,
    consider the two-generation SM to be the UV-complete theory and
    the Fermi-theory as the EFT.  Had the charm-quark been absent in
    the UV-theory but Cabibbo mixing present, the integrating out of
    the $W$-- and the $Z$--fields would not only have generated the
    usual CC and NC interactions, but also large FCNC terms in the
    EFT. The reintroduction of the $c$, before the integration,
    removes the FCNC to the lowest order, but retains it at a higher
    order, and renders the FCNC proportional to $m_c^2$. In other
    words, since a symmetry in the UV-theory had forbidden the
    generation of a particular term in the EFT, its subsequent
    generation, which could occur only through the participation of
    the light fields, bore an imprint of the masses of the light
    fields ($m_c$ in this context).}. With the attendant additional
suppression, by a factor of $\alpha_{\rm wk}\, m_{\rm light}^2 / \Lambda^2$ where
$m_{\rm light}$ is the typical mass of the SM fields, the consequent value of $\zeta$ is
small enough for our effective Lagrangian to be in consonance with
$B_s$--$\bar B_s$ mixing. In other words, this constraint should be
considered as only an indicative one, perhaps pointing to the nature of
the UV completion. 

Thus, we are brought back to the assertion implicit in the entire
discussion of this paper, namely that there has to exist some symmetry in
the UV-complete theory that ensures that the discussed operators (in a
given scenario) constitute the entire set appearing at the lowest
order in the said EFT.  The Wilson coefficients of any other
four-fermion operator generated as a result of quantum corrections
must, necessarily, be suppressed by at least $\alpha_{\rm wk}\, m_{\rm
  light}^2 / \Lambda^2$. Such a suppression, of course, would 
render the scenario safe from the perspective of $\Delta M_s$. 

Very similar to the discussion above is the
case for $\tau \to 3 \mu$, putatively generated by a quark loop.


\section{Conclusion}
\label{sec:conclusion}

In this paper, we identify the minimal extension of the SM
in terms of effective four-fermion operators that can explain two sets
of anomalies: $R_K$ and $R_{K^*}$ on the one hand, as well as $R(D)$,
$R(D^*)$ and $\rJpsi$ on the other. Explaining both sets at a single stroke has
been challenging for two reasons: $(i)$ there is a deficiency in the
former case but an excess in the latter, and $(ii)$ because they
involve different leptons, viz. muons for the first pair and $\tau$s
for the second.
 
The final state leptons, though, can be related by postulating a small
rotation of the original charged lepton field involved in the
dimension-6 operator(s). With the very inclusion of a
flavour-nonuniversal operator, such a rotation is no longer a trivial
one (as is the case with the SM). With the neutrino flavour in such
decays not being observed, only the incoherent sum over states is a
measurable quantity. As for excesses in both the neutral- and
charged-current processes, these are related by the usual $SU(2)_L$
symmetry.
 
Based on these principles, we have formulated several scenarios with a
minimal set of dimension-6 gauge and Lorentz invariant operator. The
``models'' have at most three parameters, namely the WCs corresponding
to the effective operators (two or less) and the lepton mixing angle.

Taking all the data into account, we find that two such operators are
enough to get an acceptable fit. For the best fit points, all the
observables, barring $R_{K^*}^{\rm low}$ and ${\rm BR}(B_s \to \phi \mu \mu)$ in
the low-$q^2$ bin, are consistent within $1\sigma$. (For the latter,
the agreement is better than $2 \sigma$.)  Even for the standout
observable (for which the data is still not of great quality), the
disagreement is only slightly worse than $1\sigma$. In addition, we
can also explain the observed suppression in the low-$q^2$ bins for
the decay $B_s\to \phi\mu\mu$.

A strong prediction of our analysis is that either $B\to K\mu\tau$
and/or $B_s\to\tau^+\tau^-$ will be close to discovery, and one should
look for such channels in LHCb as well as Belle-II. At the same time,
we do not attempt to probe the origin of these new operators; while a
$Z'$ or a vector leptoquark may do the job, this is left for the model
builders.

Although we start with a simplistic scenario with two operators in each
case, it is conceivable that a hitherto unknown symmetry relates the
two unknown WCs.  Indeed the choices $A_2 = A_1 / 2$, 
applicable for Models I and II, are strikingly simple, and
conceivably, may arise from some unidentified flavour dynamics. Such a
symmetry, if exact, would lead to a vanishing NP contribution (at the
tree level in the effective theory) to the $b \to s \nu \bar \nu$
amplitude. However, quantum corrections would be expected to break
this symmetry. Although such rates would be small, they should still
be visible at Belle-II.

Model I is interesting from a different perspective. The best-fit
value of the WC $A_1$ is two-orders of magnitude below the other,
namely $A_2$. Indeed, this is one case where a single operator does
almost as well as two together, 
or in other words, only two parameters for the new physics are required here. This is a
remarkably simple solution to all of the disparate set of anomalies
that confront us. A slight modified version of this case has been discussed in Ref.~\cite{our-prl}, where the bounds from $B_s\to \tau \tau$ are well under control. Most interestingly, the scale of the new (flavour)
physics is suggested to be a few TeVs at best, rendering the situation
extremely attractive for the current run of the LHC.

It has to be noted, though, that the recent LHCb bound on $B_s \to
\tau^+ \tau^-$ rules out the simplest of the scenarios. While this
measurement is crucially dependent on the use of neural networks
etc. (as the $\tau$s are yet to be fully reconstructed) and the
consequent uncertainties, it is worthwhile to investigate if the
suppression of the $C_{10}^{\rm NP}$ contribution that it calls for,
can be accommodated in such scenarios. We find that this can indeed be
done without the introduction of additional parameters, but at the
cost of introducing an additional gauge invariant operator, whose WC
is envisaged to be related with the other WCs by some
yet-to-be-discovered symmetry.  While the $\chi^2$ worsens marginally,
it is still miles better than that in the SM.  The models have a few
generic predictions, like the possibility of observing $B\to
K^{(*)}\mu\tau$ or $B_s\to\tau^+\tau^-$ in near future, and possibly
$B\to K^{(*)}$ plus missing energy. They will definitely be checked
within the next couple of years at LHCb, and Belle-II will be able to
make precision studies on these observables.  The other features of
the scenarios, including the possibility of direct observation of
TeV-scale resonances at the LHC, remain unaltered. 


The authors thank Gino Isidori and Sudhendu RaiChoudhury for
illuminating discussions. AK thanks the Science and Engineering
Research Board (SERB), Government of India, for a research grant.  DC
acknowledges partial support from the European Union's Horizon 2020
research and innovation program under Marie Sk{\l}odowska--Curie grant
  No 674896.

\appendix


\section{Appendix}
\label{sec:appendix}
The expressions for the angular coefficients present in the differential distribution of $B\to V \ell^+\ell^-$ decay, discussed in Sec.~\ref{sec:SM}, are
\begin{align}
\label{eq:I1s}
  I_1^s & = \frac{(2+\beta^2)}{4} \Big[|\apeL|^2 + |\apaL|^2 +
    (L\to R) \Big]
    +\frac{4m^2}{q^2}\Re(\apeL^{}\apeR^{*}+\apaL^{}\apaR^{*}), \\ 
\label{eq:I1c}
  I_1^c & = |\azeL|^2 \!+\!|\azeR|^2\!+\!\frac{4m^2}{q^2}\big[\abs{{\cal A}_t}^2
  \!+\!2\Re(\azeL^{}\azeR^{*})\big],\\
  I_2^s & = \frac{\beta^2}{4}\Big[ |\apeL|^2+ |\apaL|^2 + (L\to
    R)\Big],\\ 
  I_2^c & = -\beta^2 \Big[|\azeL|^2 + (L\to R)\Big].
\end{align}

Here, the transversity amplitudes $\mathcal{A}^{L,R}_{0,\|,\perp,t}$
are functions of the Wilson coefficients and the form factors
$V(q^2)$, $A_{0,1,2}(q^2)$ and $T_{1,2,3}(q^2)$ for $B\to V$
transitions. The expressions are
\begin{align}
\mathcal{A}_\perp^{L,R}&=N \sqrt{2} \lambda^{1/2} \left\{\left[ \left(C_9+ \Cnine \right) \mp \left(C_{10}+ \Cten \right) \right] \frac{V(q^2)}{m_B+m_V} +\frac{2 m_b}{q^2}C_7 T_1(q^2) \right\},\\
\mathcal{A}_\|^{L,R}&= -N \sqrt{2} \left(m_B^2 -m_V^2\right) \left\{\left[ \left(C_9+ \Cnine \right) \mp \left(C_{10}+ \Cten \right) \right] \frac{A_1(q^2)}{m_B-m_V} +\frac{2 m_b}{q^2}C_7 T_2(q^2) \right\},\\ 
\mathcal{A}_0^{L,R}&= -\frac{N}{2 m_V \sqrt{q^2}}  \Bigg\{\left[ \left(C_9+ \Cnine \right) \mp \left(C_{10}+ \Cten \right) \right] \nonumber \\ 
&\hspace*{2.cm} \times \left[\left(m_B^2-m_V^2-q^2\right) \left(m_B +m_V\right)A_1(q^2)- \lambda \frac{A_2(q^2)}{m_B+m_V}  \right] \nonumber \\
&\hspace*{2.cm} +2 m_bC_7 \left[\left(m_B^2+3 m_V^2-q^2\right)T_2(q^2)- \lambda \frac{T_3(q^2)}{m_B^2-m_V^2}  \right] \Bigg\},\\
\mathcal{A}_t&= \frac{N}{\sqrt{q^2}} \lambda^{1/2} \,2 \left(C_{10}+ \Cten \right) A_0(q^2),
\end{align}

where\hspace*{1.cm} $N= V_{tb}V_{ts}^* \Big[ \displaystyle\frac{G_F^2
    \alpha^2}{3\times 2^{10}\, \pi^5\, m_B^3\, } q^2 \,\lambda^{1/2}\,
\beta\Big]^{1/2}$,\hspace*{1.cm} with $\beta= \sqrt{1-
  4m_\mu^2/q^2}$,\\

and $\lambda \equiv \lambda(m_B^2,m_V^2,q^2)=  m_B^4 + m_V^4 + q^4 -2 (m_B^2 m_V^2 + m_V^2 q^2 + m_B^2 q^2)$.


\end{document}